\documentclass[preprint]{aastex}
\def\beq{\begin{equation}}
\def\eeq{\end{equation}}
\def\bey{\begin{eqnarray}}
\def\eey{\end{eqnarray}}

\def\mpc{\,h^{-1}{\rm {Mpc}}}
\def\kms{\,{\rm {km\, s^{-1}}}}
\def\msun{{M_\odot}}

\def\zetarr{\zeta(r_{12},r_{23},r_{31})}

\def\xir#1{\xi(r_{#1})}

\def\wrp#1{w(r_{p#1})}
\def\gs{\mathrel{\raise1.16pt\hbox{$>$}\kern-7.0pt
\lower3.06pt\hbox{{$\scriptstyle \sim$}}}}
\def\ls{\mathrel{\raise1.16pt\hbox{$<$}\kern-7.0pt
\lower3.06pt\hbox{{$\scriptstyle \sim$}}}}
\def\gtsima{$\; \buildrel > \over \sim \;$}
\def\ltsima{$\; \buildrel < \over \sim \;$}
\def\prosima{$\; \buildrel \propto \over \sim \;$}
\def\gsim{\lower.5ex\hbox{\gtsima}}
\def\lsim{\lower.5ex\hbox{\ltsima}}
\def\simgt{\lower.5ex\hbox{\gtsima}}
\def\simlt{\lower.5ex\hbox{\ltsima}}
\def\simpr{\lower.5ex\hbox{\prosima}}

\shorttitle{Galaxy clustering in the PSCz survey}
\shortauthors{Jing, B\"orner \& Suto}
\begin{document}
\title {
Spatial correlation functions and the pairwise peculiar
velocity dispersion of galaxies in the PSCz survey:
implications for the galaxy biasing in cold dark matter models
}
\author{Y.P. Jing} 
\affil{Shanghai Astronomical Observatory, the Partner Group of MPI f\"ur
Astrophysik, \\Nandan Road 80,  Shanghai 200030, China}
\email{ypjing@center.shao.ac.cn}
\author {Gerhard B\"orner} 
\affil {Max-Planck-Institut f\"ur Astrophysik,
Karl-Schwarzschild-Strasse 1, \\  85748 Garching, Germany}
\email{grb@mpa-garching.mpg.de}
%%-----------------------------------------------------------------------
\and
\author{Yasushi Suto} 
\affil{Department of Physics and Research Center for the Early Universe
(RESCEU)\\ School of Science, University of Tokyo, Tokyo 113-0033,
Japan.}
\email{suto@phys.s.u-tokyo.ac.jp}

\received{2001 March}
\accepted{2001 September }
\begin{abstract}
  We report on the measurement of the two-point correlation function
  and the pairwise peculiar velocity of galaxies in the IRAS PSCz
  survey. We compute these statistics first in redshift space, and then
  obtain the projected functions which have simple relations to the
  real-space correlation functions on the basis of the method developed
  earlier in analyzing the Las Campanas Redshift Survey (LCRS) by Jing,
  Mo, \& B\"orner (1998).  We find that the real space two-point
  correlation function can be fitted to a power law $\xi(r) =
  (r_0/r)^{\gamma}$ with $\gamma=1.69$ and $r_0=3.70 \mpc$. The pairwise
  peculiar velocity dispersion $\sigma_{12}(r_p)$ is close to $400 \kms$
  at $r_p=3\mpc$ and decreases to about $150 \kms$ at $r_p \approx 0.2
  \mpc$.  These values are significantly lower than those obtained from
  the LCRS.

 In order to understand the implications of those measurements on the
 galaxy biasing, we construct mock samples for a low density
 spatially-flat cold dark matter model ($\Omega_0 = 0.3$,
 $\lambda_0=0.7$, $\Gamma=0.2$, $\sigma_8=1$) using a set of
 high-resolution N-body simulations in a boxsize of $100h^{-1}$Mpc,
 $300h^{-1}$Mpc, and $800h^{-1}$Mpc.  Applying a stronger
 cluster-underweight biasing ($\propto M^{-0.25}$) than for the LCRS
 ($\propto M^{-0.08}$), we are able to reproduce these observational
 data, except for the strong decrease of the pairwise peculiar velocity
 at small separations. This is qualitatively ascribed to the different
 morphological mixture of galaxies in the two catalogues. Disk-dominated
 galaxy samples drawn from the theoretically constructed GIF catalog
 yield results rather similar to our mock samples with the simple
 cluster-underweight biasing. We further apply the phenomenological
 biasing model in our N-body mock samples which takes account of the
 density-morphology relation of galaxies in clusters. The model 
 does not reduce the velocity dispersions of galaxies to the level
 measured in the PSCz data either. Thus we conclude that the peculiar
 velocity dispersions of the PSCz galaxies require a biasing model which
 substantially reduces the peculiar velocity dispersion on small scales
 relative to their spatial clustering.
\end{abstract}


\section {Introduction}

Large catalogs of galaxies with their redshift and angular positions are
the major astronomical data sets which provide quantitative information
on the distribution and formation of the universe.  To properly decipher
this information requires the use of statistical tools, and to assess
their importance heavily relies on a quantitative comparison with
theoretical models.

In a recent series of papers \citep{jmb98,jb98a,jb01}, various
statistical quantities from the Las Campanas Redshift Survey (LCRS)
have been determined including the two-point correlation function
(2PCF), $\xi(r)$, the power spectrum, $P(k)$, the pairwise peculiar
velocity dispersion (PVD), $\sigma_{12}(r)$, and the three-point
correlation function (3PCF), $\zetarr$. In addition, a detailed
comparison between these observational results and the predictions of
current cold dark matter (CDM) models has been carried out. Jing, Mo,
\& B\"orner (1998; JMB98 hereafter) have constructed 60 mock samples
for each theoretical model from a large set of high-resolution N-body
simulations.The observational selection effects can then be taken into
account in the analyses of the theoretical models exactly as for the
real observational data. JMB98 have demonstrated at length that such a
procedure is essential for the proper comparison between models and
observations.

JMB98 found that both the real-space 2PCF and the PVD can be measured
reliably from the LCRS, and that the observed 2PCF for LCRS is
significantly flatter than the mass 2PCF in CDM models on scales $\ls 1
\mpc$. The observed PVD also turned out to be lower than that of the
dark matter particles in these models. JMB98 introduced a
cluster-underweight (CLW) biasing model to account for these
discrepancies; CLW essentially assumes that the number of galaxies per
unit dark matter mass in a massive halo of mass $M$ decreases as
$\propto M^{-\alpha}$.  If $\alpha=0.08$, the 2PCF and the PVD of the
LCRS are well reproduced in a spatially-flat CDM model with
$\Omega_0^{0.6}\sigma_8 \approx 0.4$, where $\Omega_0$ is the 
density parameter of
the model and  $\sigma_8$ is the current rms linear density
fluctuation within a sphere of radius $8\mpc$.  In fact, \citet{carl96}
argue that this anti-bias is at the level observed for rich clusters of
galaxies. We note here that this CLW has been applied also in recent
analytic modeling of the galaxy distribution
\citep{seljak00,pea2000,sheth00}.

In the present paper, we report on the measurements of the 2PCF and PVD
of galaxies in another large redshift survey, the PSCz catalog, which
has become publicly available recently \citep{saunders00}. Since
galaxies in the PSCz survey are selected from the IRAS point source
catalog, they are supposed to be preferentially dominated by late types
in contrast to the LCRS galaxies. Thus the difference of those
clustering statistics between LRCS and PSCz should be interpreted as an
indication for the morphology-dependent biasing of galaxies.

The rest of the paper is organized as follows; we briefly describe the
PSCz sample and the corresponding mock samples from simulations in \S 2.
Then we present the results for the 2PCF of the PSCz in \S 3 which are
compared with various model predictions in \S 4.  Also shown there are
the results for spiral {\it galaxies} predicted by the GIF simulation
\citep{gif99a,gif99b}.  Similarly we present in \S 5 and \S 6 the
observational determination of PVD and its theoretical implications,
respectively. Finally \S 7 contains a summary and further discussion.

\section{The PSCz galaxy sample and our mock catalogs}

The PSCz survey \citep{saunders00} contains the redshifts for galaxies
selected from the IRAS Point Source Catalog (PSC; Beichman et
al. 1988).  In total the PSCz catalogue contains 15411 galaxies with
their flux at $60 {\rm \mu m}$ exceeding $0.6 \rm{Jy}$ across $84\% $
of the entire sky.  The high sky coverage and the uniformity make this
publicly available survey an ideal data set for the statistical
studies of galaxy clustering.  Both those parts of the sky excluded
from the survey and the selection function are well documented in
\cite{saunders00}.  In the analysis below, we use a volume-limited
sample of galaxies with redshifts less than $6000 \kms$ according to
the flux at $60 \rm{\mu m}$ and also all the galaxies with redshifts
between $6000 \kms$ and $20000 \kms$, resulting in a final semi-volume
limited sample of $9425$ galaxies. The volume-limited sample within
$6000 \kms$ should reduce the influence of the local supercluster on
the statistics on small scales, thus minimizing possible bias from the
local supercluster. The redshift selection function is the same
as that given by \cite{saunders00} for $6000 \le z \le 20000 \kms$, is
a constant for $z< 6000 \kms$ and zero for $z> 20000 \kms$.

We construct a large set of mock samples, which have the same survey
geometry, selection function and sampling rate as the real PSCz
galaxies, from high-resolution cosmological N-body simulations.  We
emphasize that this is a very important aspect of our analysis in
asserting the statistical significance of the results.  For this
purpose, we mainly consider a spatially-flat model with ($\Omega_0$,
$\lambda_0$, $\Gamma$, $\sigma_8$)= (0.3, 0.7, 0.2, 1.0) simulated in a
comoving boxsize of $300\mpc$, where $\lambda_0$ is the cosmological
constant and $\Gamma=\Omega_0 h$ is the shape parameter of the CDM power
spectrum \citep{bbks86}.  We use three different realizations for each
model employing $256^{3} \sim$ 17 million particles
\citep{jingsuto98,jing98}.  This set of parameters is now considered as
the {\it standard} low-density CDM model. JMB98 have found that this
model reproduces the 2PCF and the PVD of the LCRS reasonably well. An
even better fit to the LCRS data was obtained for a model with $\Omega_0
= 0.2$. The changes for this model can be roughly seen by a simple
scaling of the term $\sigma_8 \Omega_0^{0.6}$ which is proportional to
the amplitude of the PVD. For reference, we also consider a
$\lambda_0=0$ {\it Standard} CDM model simulated in a $300\mpc$ box as
well as additional LCDM models in $100h^{-1}$Mpc and $800h^{-1}$Mpc to
make sure that our conclusions are free from the numerical
artifacts. The parameters for those simulation models are summarized in
Table~\ref{tab:simulation}.

The mock samples are generated in the same way as described in JMB98,
using the selection function from the PSCz \citep{saunders00}. We
construct 20 mock samples for each realization, and thus altogether 60
mock samples for the LCDM model in a $300\mpc$ box . For each mock
sample we apply the CLW bias (JMB98). The CLW model, by construction,
gives less weight to the denser clusters than the dark matter model,
and thus is especially appropriate for the PSCz which is supposed to
be dominated by spiral galaxies preferentially avoiding the
high-density cluster regions.  To be more specific, we randomly select
{\it galaxies} from the dark matter particles so that the number of
galaxies per unit dark matter mass, $N/M$, becomes proportional to
$M^{-\alpha}$ within a halo of mass $M$.  We apply this anti-biasing
scheme for all dark halos with $M > 7 \times 10^{11}h^{-1}\msun$.
While the model with $\alpha=0.08$ reproduces the 2PCF and the PVD for
the LCRS data, we find (see below) that the PSCz data require
${\alpha} = 0.25$, indicating stronger anti-biasing than the optically
selected galaxies. We also apply this same bias model to the SCDM when
we compare the predictions of this model to the observations.

In addition, we also construct 20 mock catalogs from the GIF data base
\citep{gif99a,gif99b} for comparison to our models and to the data. The
GIF simulation adopts the same cosmological model, but with a slightly
lower normalization of the linear power spectrum $\sigma_8=0.9$.  The
simulation box size is $141\mpc$. The most important feature of the GIF
data is that a semi-analytic galaxy formation model is implemented in
contrast to our biasing scheme entirely on the basis of the
gravitational clustering.  We select ``spiral galaxies'' from the GIF
mock samples according to their bulge-to-total luminosity ratio $B/T$.
About 80\% of those mock galaxies at the low end of $B/T$ are retained
for our analysis, taking into consideration the fact that the population
of ellipticals is about $1/4$ that of spirals in the local Universe.

It might be more appropriate to select ``galaxies'' according to their
infrared fluxes for the present purpose. Fortunately, however, the
clustering properties of IRAS galaxies do not sensitively depend on the
infrared luminosity \citep{szapudi}, nor do those of the GIF spirals
seem to depend sensitively on the ratio $B/T$. Thus our results below
are fairly robust with respect to small changes in the
population of galaxies.

\section{The two-point correlation functions for the PSCz galaxies}

Quite a number of large galaxy catalogs, both two-dimensional (angular),
and three-dimensional (redshift) catalogs, have been used to determine
the 2PCF, and this statistic has now been established quite well. This
statistic has produced constraints on theoretical models which are
sensitive to the cosmological parameters, the initial power spectrum of
the mass density fluctuations, and the biasing of specific galaxy types
with respect to the dark matter, among others.

We use the following estimator to measure the 2PCF $\xi_z$ in redshift
space;
%%%%%%%%%%%%%%%%%%%%%%%%%%%%%%%%%%%%%%%%%%%%%%%%%%%%%%%%%%%%%%
\bey\label{xizest}
\xi_z(r_p,\pi)&=&{RR(r_p,\pi) \times DD(r_p,\pi) \over
DR(r_p,\pi)^2}-1 , 
\eey
%%%%%%%%%%%%%%%%%%%%%%%%%%%%%%%%%%%%%%%%%%%%%%%%%%%%%%%%%%%%%%
where DD is the count of galaxy-galaxy pairs in the projected separation
$r_p$ and radial separation $\pi$ bins, RR and DR are similar counts of
pairs formed by two random points and by one galaxy and one random
point, respectively.

\begin{figure}
\epsscale{1.0} \plotone{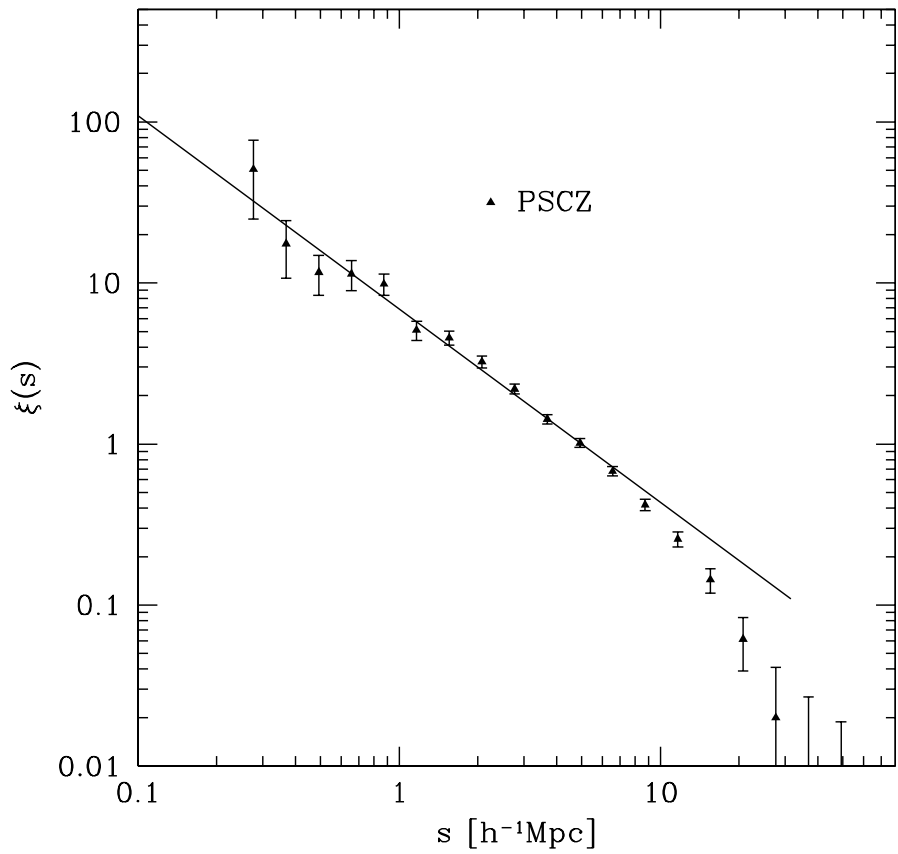}
\caption{The redshift two-point correlation function measured from the
PSCz catalog (filled triangles). The straight line is a power-law 
$(5\mpc/s)^{1.2}$.
}\label{fig1}\end{figure}

We first display the redshift correlation function $\xi(s)$ in
Figure~\ref{fig1}, which is the circular average of $\xi_z(r_p,\pi)$
for a radius $s=\sqrt{r_p^2+\pi^2}$.  The error bars have been
estimated by the bootstrap resampling technique using the formula of
\citet{mo92}. The resulting 2PCF is well fitted by
$\xi(s)=(5\mpc/s)^{1.2}$ for $s\le 8\mpc$, but falls faster than the
power-law on larger separations in good agreement with previous
results \citep{seaborne99}. Here we focus our discussion on the
projected 2PCF and the PVD \citep{dp83}, not only because these two
statistics have not been determined for the PSCz galaxies before as of
the writing of this paper but also because they provide very useful
independent information about the spatial and velocity distribution of
IRAS galaxies which can be used to understand and test galaxy
formation models.

\begin{figure}
\epsscale{1.0} \plotone{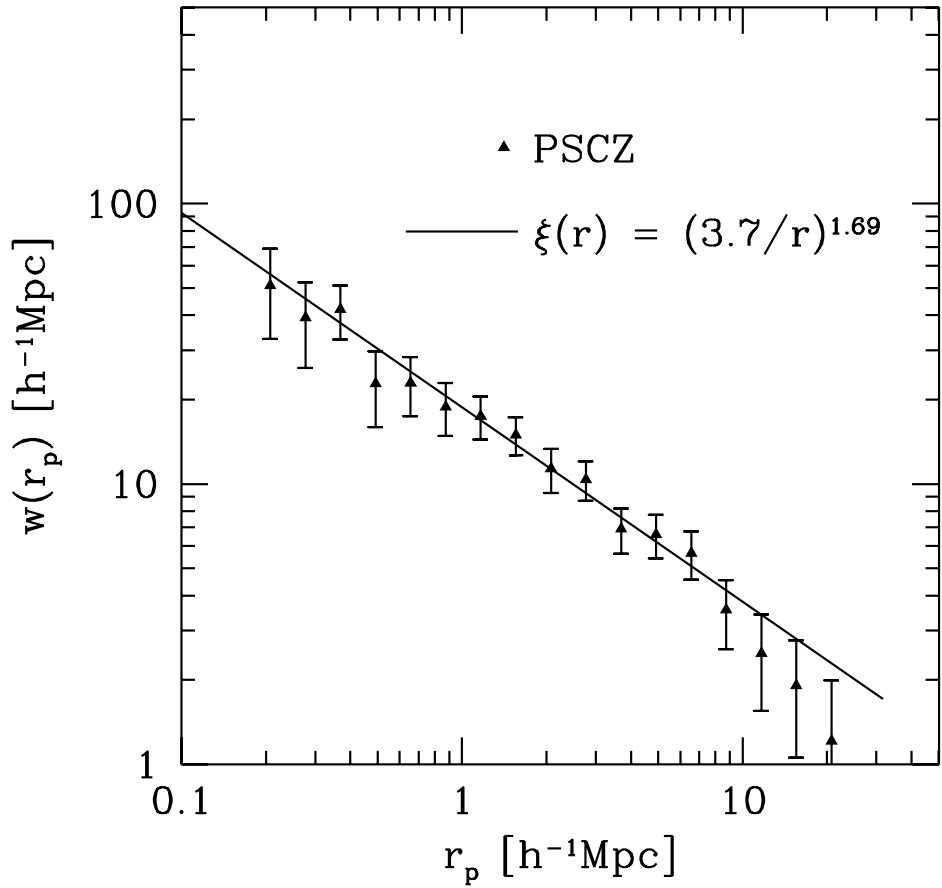}
\caption{
The projected two-point correlation function measured from the
PSCz catalog (filled triangles). Error bars are
  $1\sigma$ deviations given by bootstrap resampling. 
The solid line is the best power-law fit.
  }\label{fig2}\end{figure}

We estimate the projected 2PCF $w(r_p)$ from the integrated redshift
2PCF as\footnote{Note that our current definition of $w(r_p)$ is a
factor of two smaller than that of \cite{dp83}, for example.}
%%%%%%%%%%%%%%%%%%%%%%%%%%%%%%%%%%%%%%%%%%%%%%%%%%%%%%%%%%%%%%%%%%%
\beq\label{wrp}
w(r_p)=\int_0^\infty\xi_z(r_p,\pi)d\pi =\sum_{i}\xi_z(r_p,\pi_{i})
\Delta\pi_{i} ,
\eeq
%%%%%%%%%%%%%%%%%%%%%%%%%%%%%%%%%%%%%%%%%%%%%%%%%%%%%%%%%%%%%%%%%%%
where the summation runs from $\pi_{1}=0.5 \mpc$ to $\pi_{50}=49.5 \mpc$
with $\Delta \pi_{i} = 1 \mpc$.  JMB98 have shown that this method
yields an unbiased estimate of $w(r_p)$ which is also related to the
real space 2PCF simply as
%%%%%%%%%%%%%%%%%%%%%%%%%%%%%%%%%%%%%%%%%%%%%%%%
\beq\label{wxi}
w(r_p)=\int_0^\infty \xi(\sqrt{r_p^2+y^2}) dy .
\eeq
%%%%%%%%%%%%%%%%%%%%%%%%%%%%%%%%%%%%%%%%%%%%%%%%
Figure~\ref{fig2} plot the projected 2PCF for the PSCz with error bars
estimated again with the bootstrap resampling technique.  Our tests with
the mock samples indicated that a standard scatter among different mock
samples is actually comparable to these bootstrap errors.  So we will
use the bootstrap error for the observational estimates and the mock
sample scatter for the comparison of the models with the observations.

If the real space 2PCF $\xir{}$ is approximated as a power-law of the
form:
%%%%%%%%%%%%%%%%%%%%%%%%%%%%%%%%%%%%%%
\beq\label{xir}
         \xi(r) = (r_0/r)^\gamma  ,
\eeq
%%%%%%%%%%%%%%%%%%%%%%%%%%%%%%%%%%%%%%
then the projected 2PCF $\wrp{}$ reduces to
%%%%%%%%%%%%%%%%%%%%%%%%%%%%%%%%%%%%%%%%%%%%%%%%%%%%%%%%%%%%%%%%%%%%%
\beq
   \wrp{} = \sqrt{\pi} 
\frac{\Gamma(\gamma/2-1/2)}{\Gamma(\gamma/2)}(r_0/r_p)^\gamma r_p  ,
\eeq
%%%%%%%%%%%%%%%%%%%%%%%%%%%%%%%%%%%%%%%%%%%%%%%%%%%%%%%%%%%%%%%%%%%%%
with $\Gamma(x)$ being the Gamma function. As Figure \ref{fig2}
indicates, the observed $\wrp{}$ of the PCSz galaxies for $r_p \ls
20\mpc$ is fitted to the above form reasonably well with
%%%%%%%%%%%%%%%%%%%%%%%%%%%%%%%%%%%%%%%%%%%%%%%
\beq\label{plfit2}
r_0 = 3.70 \mpc ,  \qquad   \gamma =1.69  .
\eeq
%%%%%%%%%%%%%%%%%%%%%%%%%%%%%%%%%%%%%%%%%%%%%%%
The slope is shallower and the amplitude is lower than those for the
LCRS (JMB98). The error-bars are, however, substantially larger than
for the LCRS, indicating that this catalog is somewhat noisier. The
reason might be that the intrinsic clustering is weaker in this catalog
and also that the volume surveyed by the catalog is smaller and in fact
sparsely sampled. There is some indication for a deviation from the
power-law between $0.2 \mpc$ and $0.8 \mpc$, but it is not highly
significant when considering the error-bars. The above best-fit
parameters are in very good agreement with the result of the $1.2
\rm{Jy}$ sample, $\gamma=1.66$ and $r_0=3.76 \mpc$ \citep{fisher94a}.

\section{Comparison to the two-point correlation functions from the 
mock samples}
 
\begin{figure}
\epsscale{1.0} \plotone{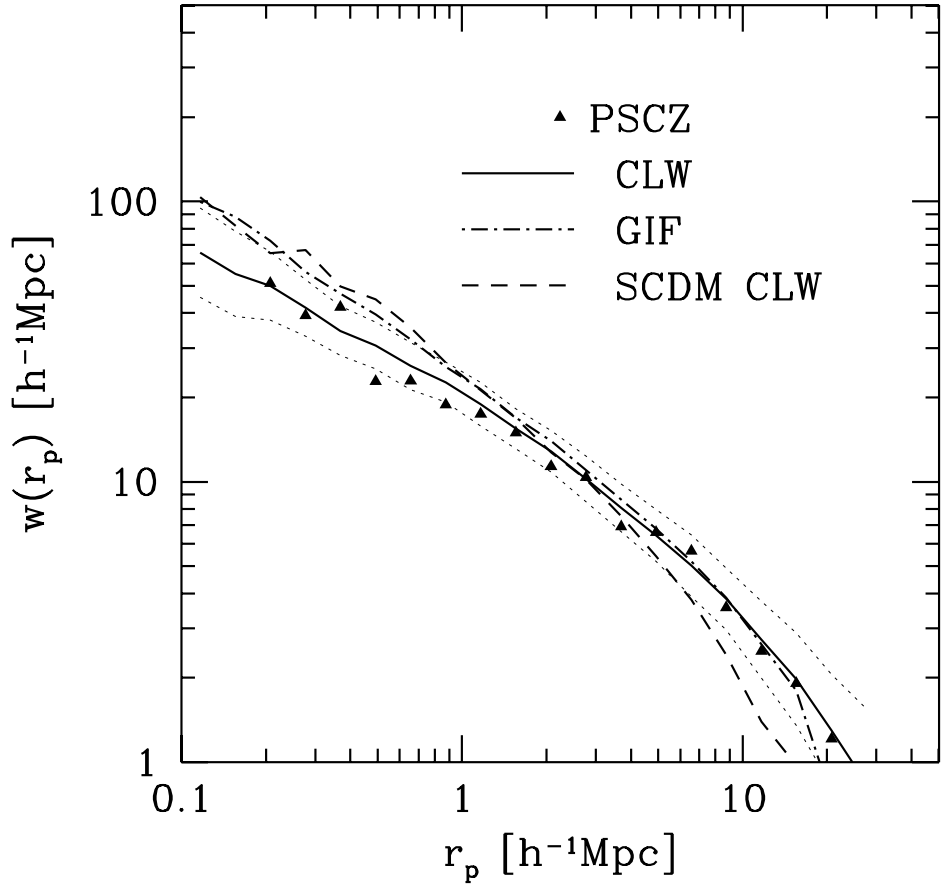}
%\epsscale{1.0} \plotone{modobs_xi.eps}
\caption{ The predictions of CDM models vs the observation for the
projected two-point correlation function. Triangles show the
observational result. The mean value and the $1\sigma$ limits
predicted by the cluster-weighted bias model are shown by the thick
and thin lines respectively, and the mean values of the GIF simulation
and the SCDM CLW model by the dot-dashed and dashed lines (without
error bars). The SCDM curve is shifted vertically by a factor of
$1/\sigma_8^2$ to account for the necessary linear bias in this model.  }
\label{fig3}\end{figure}

Figure~\ref{fig3} compares the projected 2PCF $\wrp{}$ for the PSCz
galaxies with those for the mock samples.  The CLW bias model with
$\alpha = 0.25$ reproduces the PSCz data quite well within the $1
\sigma$ error (dotted lines); even the wiggly structure of the PSCz
$\wrp{}$ below $0.8 \mpc$ is recovered.  The value of $\alpha$ larger
than for the optically selected galaxies (e.g., $\alpha=0.08$ for the
LCRS) is consistent with the observational density-morphology
relation; spiral galaxies preferentially selected by the IRAS PSCz do
not concentrate in dense cluster regions. Moreover it is interesting
that our CLW bias model can account for this behavior in a
quantitative manner, purely by an appropriate adjustment of the
parameter $\alpha$. The SCDM model, when the CLW bias is applied,
appears to underpredict clustering on scales of about $10\mpc$, as
many previous studies have already found.

To follow up this point a bit more, we have also constructed mock
catalogs from the GIF simulation data \citep{gif99a,gif99b}.  From the
simulated catalogue, we select those galaxies with $\Delta V_{bg} \equiv
V_{b}-V_{g} \ge 1$, where $V_{b}$ and $V_{g}$ denote the V-band
magnitudes of the bulge and the whole galaxy.  This encompasses about 80
percent of the galaxies in the GIF catalog. The resulting $\wrp{}$ again
fits the observations quite well (Fig.\ref{fig3}) for $r_p \simgt
1\mpc$. While the amplitude of $\wrp{}$ for the GIF simulation is a bit
larger than the PSCz data for $r_p \simlt 1\mpc$, it still lies close to
the $+1 \sigma$ error line of our CLW mock samples.  So these
semi-analytic models of galaxy formation which incorporate physical
processes like star formation and supernova explosions in some global
way have a similar effect of reducing the number of galaxies per unit
dark matter mass as our simple bias prescription.

\begin{figure}
\epsscale{1.0} \plotone{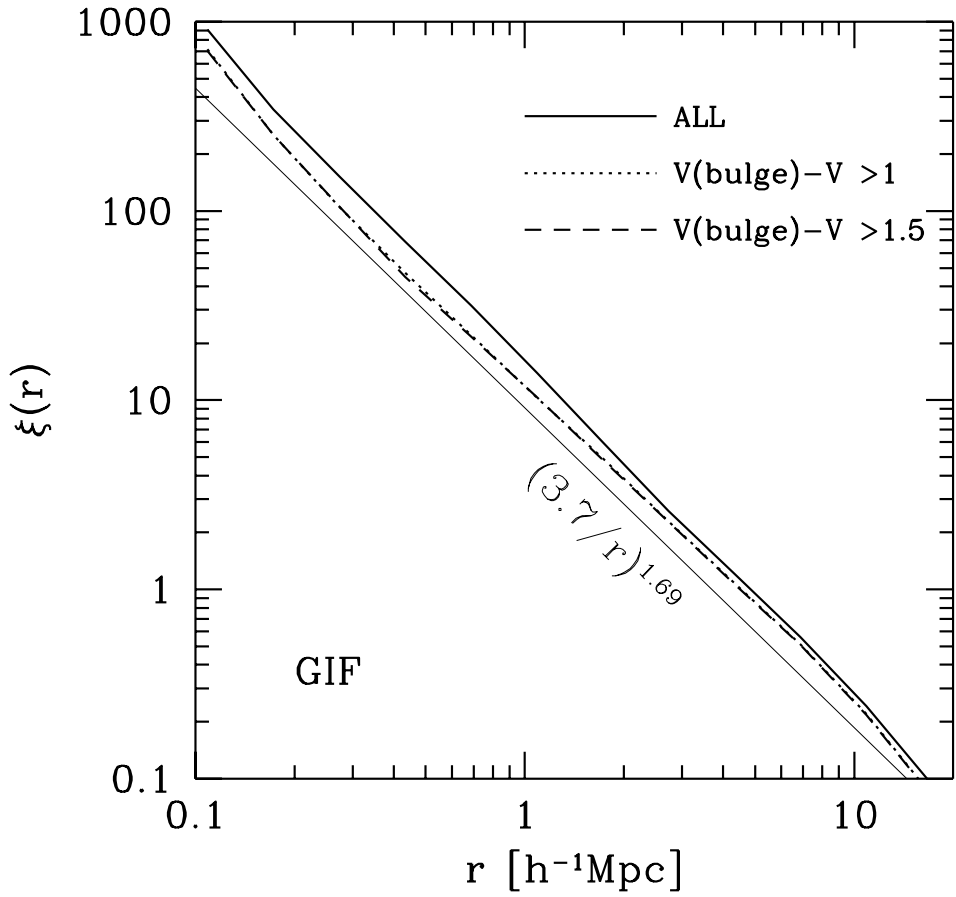}
\caption{The real space two-point correlation functions of the
galaxies selected in the GIF simulation based on the magnitude
difference $\Delta V_{bg}$ between bulge and whole galaxy. The thin 
line is the fit $\xi(r)=(3.7/r)^{1.69}$ of the IRAS galaxies.}
\label{fig4}\end{figure}

Figure~\ref{fig4} plots the 2PCFs in real space, $\xir{}$, of the GIF
mock samples with different selection criteria, all of which follow a
power-law $\propto r^{-1.69}$ quite nicely.  While selecting galaxies on
the basis of their $\Delta V_{bg}$ reduces the 2PCF at $r \simlt 1\mpc$
compared with those for all galaxies, the two different criteria,
$\Delta V_{bg} \ge 1$ and $\Delta V_{bg} \ge 1.5$, result in almost
identical 2PCFs despite the fact that the latter case selects only $60
\%$ of the entire GIF galaxies.

\begin{figure}
\epsscale{1.0} \plotone{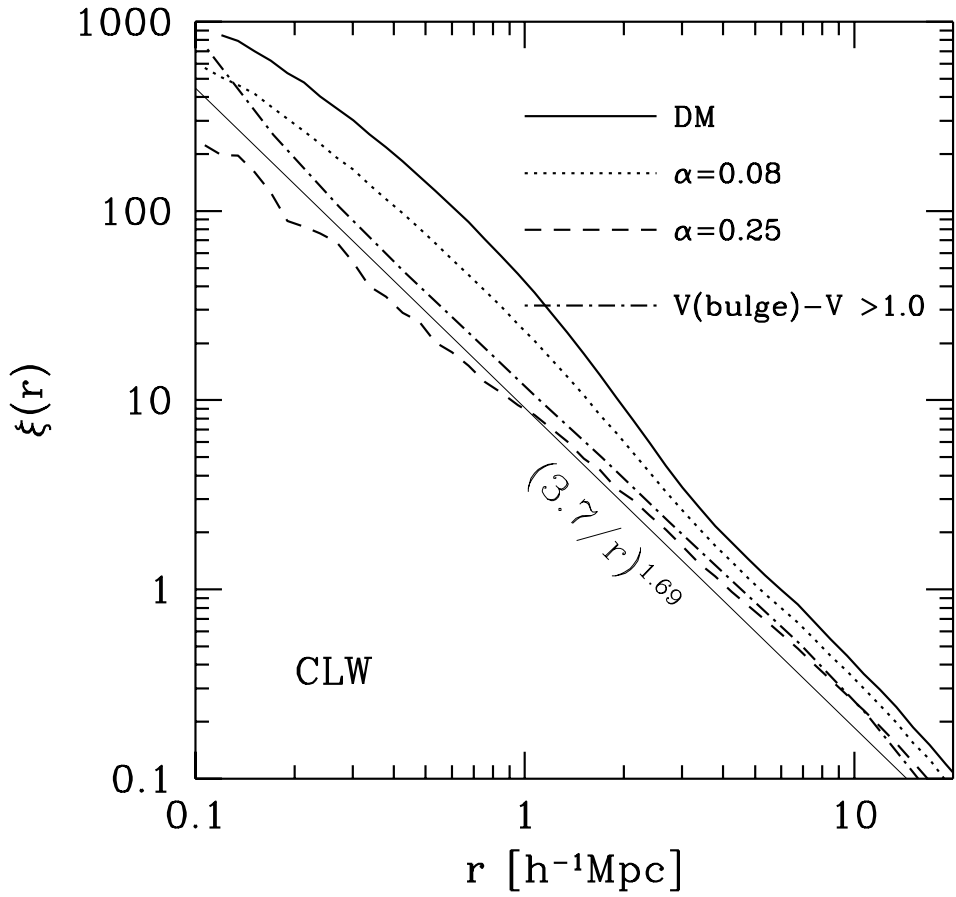}
\caption{The real space two-point correlation functions of different
bias tracers defined by the cluster-weighted model. For comparison,
the result of the galaxies with $\Delta V_{bg}>1$ in the GIF
simulation are also plotted. The thin line is the fit
$\xi(r)=(3.7/r)^{1.69}$ of the IRAS
galaxies. }\label{fig5}\end{figure}

In Figure~\ref{fig5} we display several 2PCFs for dark matter from our
simulations, the CLW mock samples with $\alpha=0.08$ and
$\alpha=0.25$, and the $\Delta V_{bg} \ge 1.0$ sample from the GIF
data set.  Clearly, as has been noted many times already (see
e.g. JMB98), the dark matter particles from the full simulation
produce a $\xir{}$ which is too large and too steep for $r \ls 1
\mpc$. As for the LCRS galaxies, a scale-dependent bias is required
for the LCDM model to be reconciled with the observed 2PCF of the PSCz
galaxies, which is consistent with the result obtained by Hamilton \&
Tegmark (2001) from an analysis of the real space power spectrum for
the PSCz galaxies. The CLW bias model with $\alpha=0.08$ that fits the
LCRS data well still predicts a much higher amplitude of $\xir{}$ than
the PSCz data which are in good agreement with $\alpha = 0.25$ model.
The GIF spiral result lies between these two CLW bias models, and is
closer to that of $\alpha = 0.25$ at $r \simgt 1 \mpc$.  We also note
that the result of all GIF galaxies is quite close to the CLW model
with $\alpha=0.08$.

In summary, we may conclude that one can reproduce the 2PCFs both for
the PSCz galaxies and for the optically selected LCRS galaxies in an
LCDM model adopting a simple biasing prescription dependent on the type
of galaxies in the sample, as long as the bias underweights galaxies in
dense clusters. The semi-analytic GIF models may be a step on the way to
provide a physical interpretation to the phenomenological CLW bias
model.

\section{The pairwise peculiar velocity dispersion for the PSCz galaxies}

The Pairwise Velocity Dispersion (PVD) of galaxies is a well-defined
statistical quantity which contains interesting information in
principle on the cosmic matter distribution.  The peculiar velocities
of galaxies are determined by the action of the local gravitational
fields, and thus they directly mirror the gravitational potentials
caused by dark and luminous matter (Juszkiewicz et al. 2000 and
references therein). The PVD is measured by modeling the distortions
in the observed redshift-space correlation function $\xi_z(r_p, \pi)$.
The basic step in modeling is to write $\xi_z(r_p, \pi)$ as a
convolution of the real-space 2PCF function $\xi(r)$ and the
distribution function $f(v_{12})$ of the relative peculiar velocity
$v_{12}$ of galaxy pairs along the line of sight \citep{dp83}:
%%%%%%%%%%%%%%%%%%%%%%%%%%%%%%%%%%%%%%%%%%%%%%%%%%%%%%%
\beq\label{xizmodel}
1+\xi_z(r_p, \pi)=
\int
f(v_{12})\left[1+\xi(\sqrt{r_p^2+(\pi-v_{12})^2})
\right]dv_{12} .
\eeq
%%%%%%%%%%%%%%%%%%%%%%%%%%%%%%%%%%%%%%%%%%%%%%%%%%%%%%%
While the real-space 2PCF $\xi(r)$ can be directly estimated by
inverting equation (\ref{wxi}), we use the power-law fit (\ref{plfit2})
as in most previous work. For the velocity distribution function,
we use the following exponential form which is 
supported from observations \citep{dp83,fisher94b},
theoretical models  \citep{diaferio96,sheth96,seto98}, and
direct simulations \citep{efstathiou,magira}:
%%%%%%%%%%%%%%%%%%%%%%%%%%%%%%%%%%%%%%%%%%%%%%%%%%%%%%%%%%%%%%%%%
\beq\label{fv12}
f(v_{12})={1\over \sqrt{2}\sigma_{12}} \exp \left(-{\sqrt{2}\over
\sigma_{12}} \left| v_{12}-\overline{v_{12}}\right| \right), 
\eeq
%%%%%%%%%%%%%%%%%%%%%%%%%%%%%%%%%%%%%%%%%%%%%%%%%%%%%%%%%%%%%%%%%
where $\overline{v_{12}}$ is the mean infall velocity and $\sigma_{12}$
is the dispersion of the 1-D pairwise peculiar velocities along the line
of sight.

It is worth pointing out that the above modeling (eqs. [\ref{xizmodel}]
and [\ref{fv12}]) is an approximation and the infall velocity
$\overline{v_{12}}$ is not known {\it a priori}, either.  JMB98
demonstrated with the LCRS mock samples that the above procedure
provides an accurate estimate of $\sigma_{12}$ (within 20\% accuracy) if
$\overline{v_{12}}$ is known separately. In other words,
$\overline{v_{12}}$ must be modeled carefully to allow a precise
measurement of the dispersion $\sigma_{12}(r)$.  In reality, however,
$\overline{v_{12}}$ in the real Universe is not fairly certain at the
present.  One might think that on small scales $\overline{v_{12}}$ is
negligible, but this is true only for very small scales indeed. The
function $\overline{v_{12}}$ rises quite sharply around $1\mpc$
\citep{mo97}, reaching twice the Hubble velocity just beyond
$1\mpc$.  Therefore the proper modeling of the infall velocity is
crucial in an accurate measurement of the PVD $\sigma_{12}(r)$.

\begin{figure}
\epsscale{1.0} \plotone{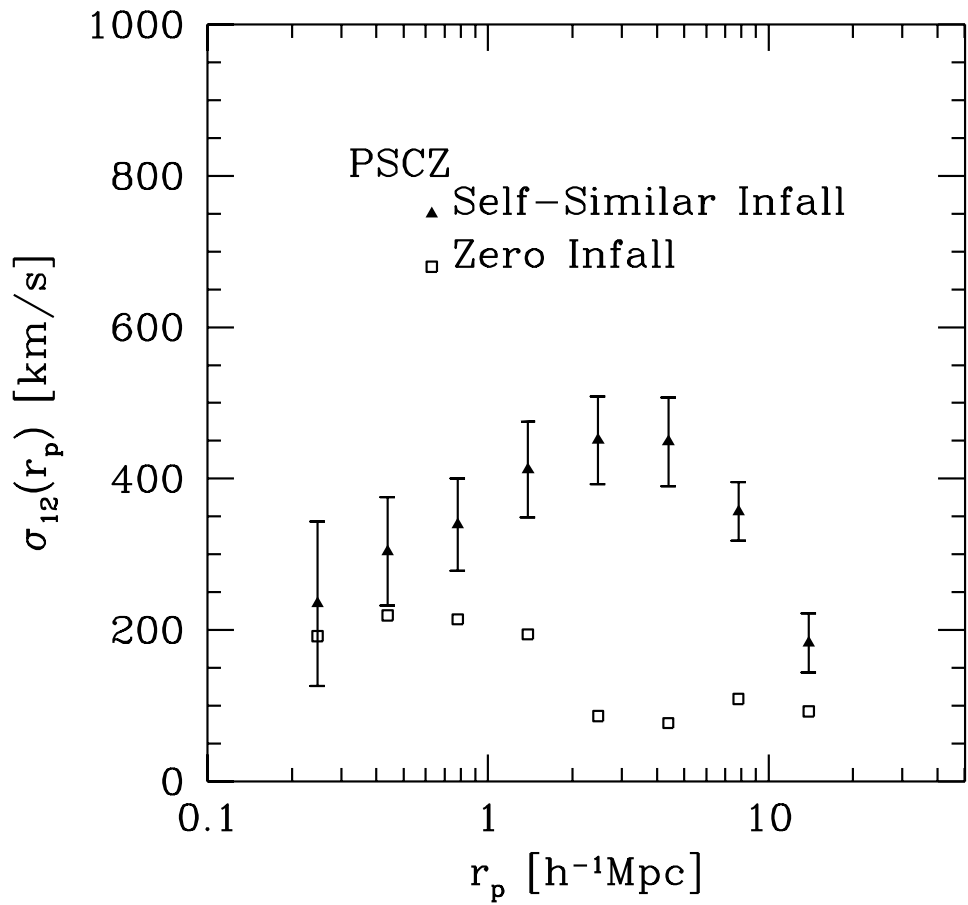}
\caption{The pairwise velocity dispersion of the
PSCz catalog determined for the self-similar infall (filled triangles) and for zero infall (open triangles).}\label{fig6}\end{figure}

Here we adopt the approach of JMB98 to determine the PVD for the
PSCz catalogue.  Figure~\ref{fig6} displays the result of this modeling
both for the case of a self-similar infall and for zero infall.  The
self-similar infall model has the same parameters as those used in
JMB98, and is quite close to the mean pair velocity in CDM models with
$\Omega_0^{0.6}\sigma_8=0.4$ for the relevant scales $r\ls 5\mpc$. As
emphasized in JMB98, only the use of some reasonable infall model
(although the result is not very sensitive to the specific model used)
gives a reliable reconstruction of the PVD, whereas the zero infall case
does not even qualitatively reproduce the true value. We therefore
consider the result of the self-similar infall as a reasonable estimate
for the pairwise velocity dispersion of the PSCz galaxies.  

The procedure yields much lower PVD for the PSCz than that for the LCRS:
$\sigma_{12}$ just reaches $300 \kms$ at $r_p = 1 \mpc$, whereas
$\sigma_{12}(1 \mpc) = 570 \pm 80 \kms$ for the LCRS (JMB98). This is
again qualitatively consistent with the fact that spirals have smaller
random motions than the galaxies that reside in big clusters.  Our
result at $r_p\approx 1\mpc$ is in good agreement with the value,
$317\kms$, of Fisher et al. (1994b) for the $1.2 \rm{Jy}$ sample,
although they did not show the scale-dependence of the PVD.

Very remarkable in Figure \ref{fig6} is the decrease of the PVD for
small separations.  The bootstrap errors in Figure \ref{fig6} are quite
substantial. Again we find that the errors are much larger than for the
LCRS.  Nevertheless it seems significant that the PVD decreases down to
about $200 \kms$ near $r_p=0.2\mpc$. In contrast the LCRS data did not
exhibit such a fall-off, but rather the PVD stayed almost constant near
$400 \kms$.  If we further take into account the observational error of
redshifts (typically $\sim 120\kms$) and add it to the PVD measurement
in quadrature, the real value becomes even smaller, $\sim 150\kms$ at
$r_p=0.2\mpc$. The galaxy pairs in the PSCz catalog are seemingly ``very
cold''; they show very little random motion.

\section{ Comparison with the pairwise velocity dispersion from the 
mock samples}

\begin{figure}
\epsscale{1.0} \plotone{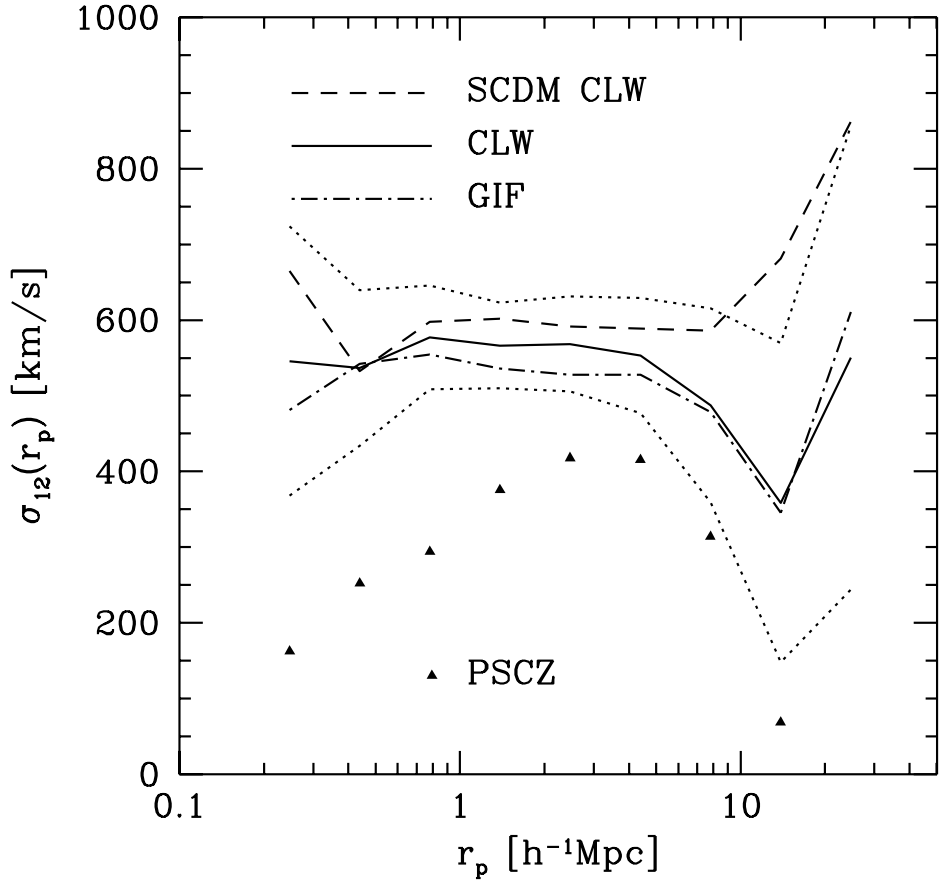} 
%\epsscale{1.0} \plotone{modobs_sig.eps} 
\caption{ The predictions of CDM models vs the observation for the
pairwise velocity dispersion. Triangles show the observational result
which has been corrected for the observational error of redshifts in
quadrature. The mean value and the $1\sigma$ limits predicted by the
cluster-weighted bias model are shown by the thick and thin lines
respectively, 
and the mean values of the GIF simulation and the SCDM CLW
model by the dot-dashed and dashed lines (without error bars). }
\label{fig7}\end{figure}

Figure~\ref{fig7} compares the PVD for the CLW and GIF mock samples
with that for the PSCz galaxies. In all cases, the model predictions
significantly exceed the estimate for the PSCz.  Only around $r_p
\approx 3 \mpc$ the disagreement is not serious, especially if the
large error bars are taken into account.  In fact, we may speculate
that a CDM model with $\Omega_0 = 0.2$ may even produce quite a good
fit to the data at larger $r_p$, since the amplitude of the PVD scales
with $\sigma_8 \Omega_0^{0.6}$. Reducing the values of $\sigma_{12}$
accordingly, brings agreement on scales larger than $3 \mpc$. There
is, however, no way to reproduce the steep decrease towards small
values ($\sim 150 \kms$) at $r_p = 0.2 \mpc$ of the PSCz data. The
pairwise velocity dispersions are rather similar in the SCDM and in
the LCDM models.

This is really puzzling, because the LCRS galaxies do not show a
signature like this. Even taking into account the large error bars at
small scales, we have to concede a formal $3 \sigma$ deviation from
the model results. This is consistent with previous findings of
Ostriker \& Suto (1990), \citet{localgroup} and
\citet{straussetal1998}, who found that the galaxies in the local
neighborhood are very quiet and that the dispersion of galaxies in
mild density regions are small compared with the CDM predictions (also
see Peebles 1995; Nolthenius, Klypin, \& Primack 1997; Baker, Davis,
\& Lin 2000).  It may also be that the measurements as well as the
models become somewhat unreliable on such small scales.  The finite
sizes of galaxies may also reduce the PVD somewhat.  As shown in
\citet{sutojing97} this can amount to up to 10 percent, if $0.1 \mpc$
is taken as a fiducial size, but the effect can be more pronounced for
large and interacting spiral galaxies.

Nonetheless our results present a new constraint on the well-defined PVD
statistic that any successful galaxy formation theory should explain. At
the moment, we may speculate that perhaps the spiral galaxies obey a
very special velocity bias, or that the CDM models are way off in the
description of small scale clustering as recently discussed in a
different context \citep{ss00,yoshida}.

\begin{figure}
\epsscale{1.0} \plotone{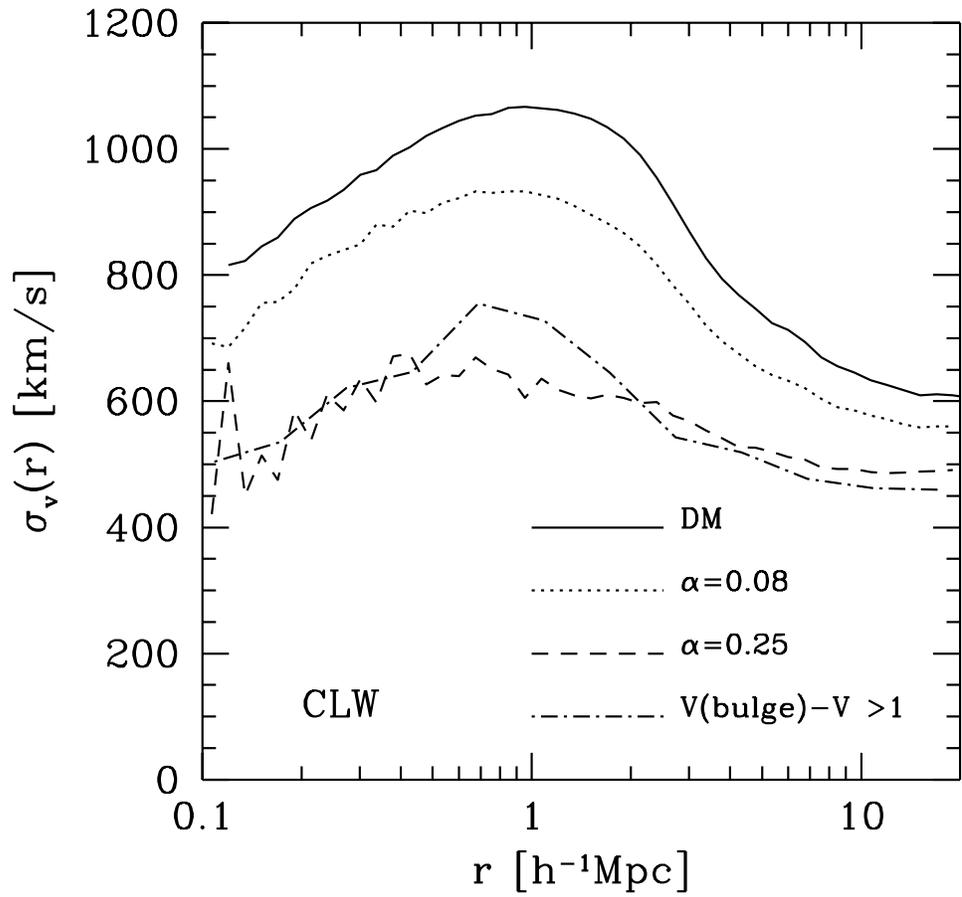} 
\caption{The pairwise velocity dispersion as determined from
the 3-D velocity, of different bias tracers defined by the
cluster-weighted model. For comparison, the result of the galaxies
with $\Delta V_{bg}>1$ in the GIF simulation are also plotted. }
\label{fig8}\end{figure}

\begin{figure}
\epsscale{1.0} \plotone{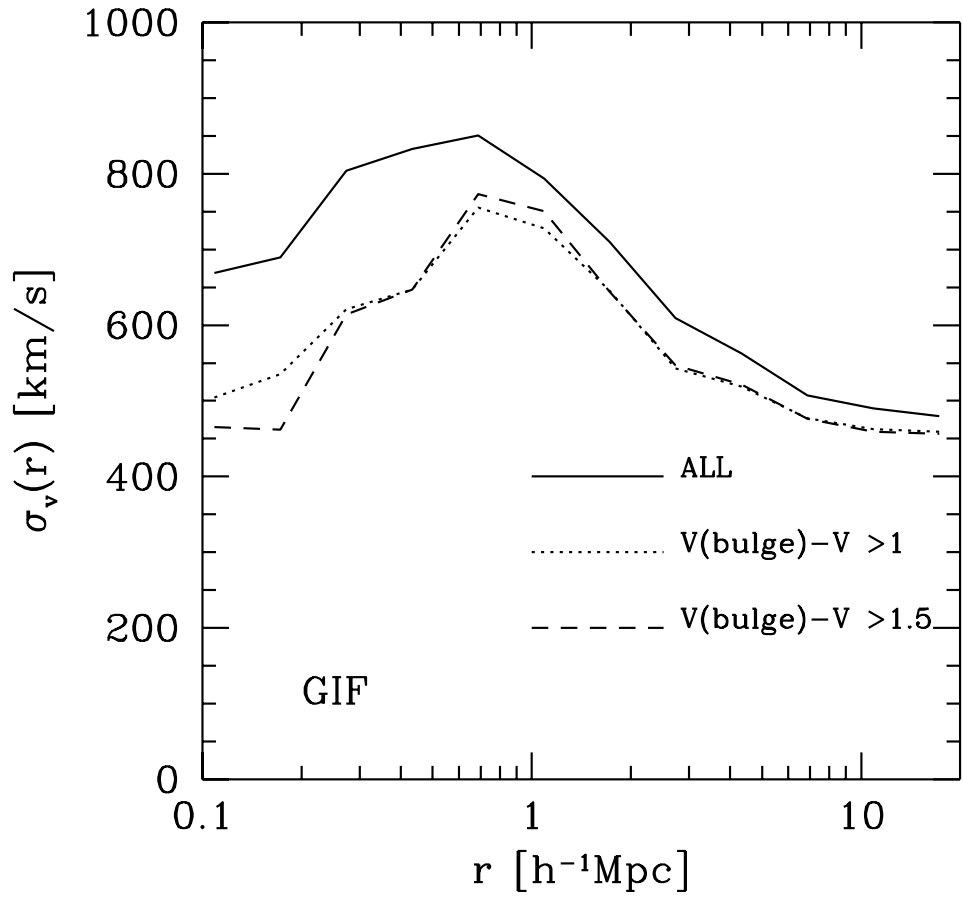} 
\caption{The pairwise velocity dispersion as determined from
the 3-D velocity, of the galaxies selected in the GIF simulation based
on the magnitude difference $\Delta V_{bg}$ between bulge and whole galaxy.}
\label{fig9}\end{figure}

To separate clearly the effect of introducing the CLW bias and of
constructing the mock catalogs, we present in Figures \ref{fig8} and
\ref{fig9} the PVD from our full simulation for the CDM model and for
the GIF simulation.  The CDM models adopting the CLW bias with $\alpha =
0.08$ and $\alpha = 0.25$ have a significantly lower PVD
than the dark matter.  We can also see that the shape of the PVD is
quite similar among the dark matter, the CLW, and the GIF samples, while the
amplitude of the dark matter particle PVD is typically higher by a few
$100 \kms$. Again the GIF galaxies with $\Delta V_{bg} \ge 1.0$ or
$\Delta V_{bg} \ge 1.5$ do not show any significant difference. The
maximum in the PVD near $1\mpc$ present in the full simulation and the
GIF sample is much reduced by the construction of the mock catalogs, and
also for the $\alpha = 0.25$ CLW models.

\begin{figure}
\epsscale{1.0} \plotone{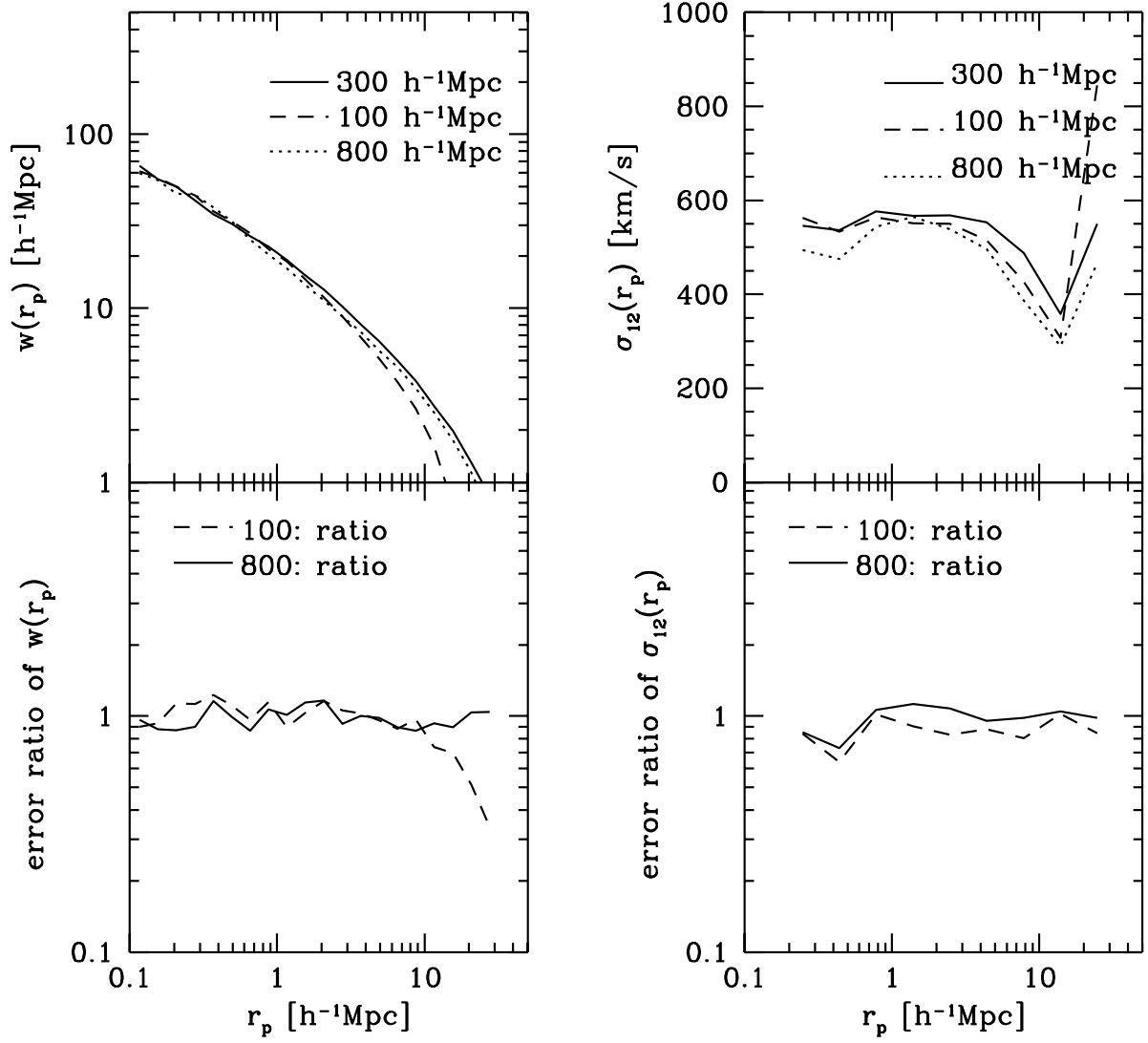} 
\caption{Effect of boxsize in numerical simulations.}
\label{fig10}\end{figure}

So far we have used the $300\mpc$ simulations alone in the comparison.
Since the boxsize is roughly the same with the PSCz survey itself once
the outer boundary of the survey is concerned (in fact the weight of
the most distant galaxies is small for our statistics and therefore
the effective volume of the survey is much smaller than as the outer
boundary indicates), our results may not faithfully reproduce the real
cosmic variance even though we have three independent realizations. On
the other hand, the small-scale peculiar velocity dispersions are
supposed to be sensitive to the numerical resolution, and the
$300\mpc$ boxsize may be too large in this sense. To show that our
$300\mpc$ simulations are not statistically biased, we compute the
projected 2PCF and the pairwise peculiar velocity dispersions for mock
samples from $100\mpc$ and $800\mpc$ LCDM simulations also (again
three realizations for each). The upper two panels of Figure
\ref{fig10} show the mean values of these two statistics. The
agreement among the mock samples from different simulation boxes is
satisfactory; the small difference between different simulations can
be easily accounted for by the simulation parameters like the large
softening length in the $800\mpc$. More important is the question if
the error measurements from the $300\mpc$ simulations are
underestimated. We show the ratio of the $1\sigma$ error from the
$800\mpc$ ($100\mpc$) simulations to that from the $300\mpc$
simulations in the lower panels. The error ratios for both statistics
and for both boxsizes are very close to 1 for the scales $\ls
10\mpc$. This comparison clearly illustrates that the expected values
and their sample-to-sample variations for $\wrp{}$ and $\sigma_{12}$
are well represented in the three realizations in the $300\mpc$
simulations and in the GIF mock samples.

\section{Discussion}

We have analyzed the data set of the IRAS PSCz galaxies
\citep{saunders00}, and computed the two-point correlation functions and
the pairwise peculiar velocity dispersion for these galaxies. 
A power-law fit to the real-space 2PCF $\xi(r) =(r_0/r)^\gamma$ gives an
exponent $\gamma=1.69$, and a lower amplitude $r_0=3.70 \mpc$ than the
value obtained
for the Las Campanas Redshift Survey ($r_0 = 5 \mpc$). We show that
these results can be very well reproduced from mock samples constructed
for a high resolution CDM simulation with $\Omega_0 = 0.3$ and
$\lambda_0 = 0.7$ if we apply a biasing model which strongly
underweights galaxies in the dense cluster regions. While this CLW bias
was originally introduced by JMB98 to fit the LCRS galaxies, the bias
needed for the PSCz galaxies is much stronger. More specifically, we
find that the number of galaxies per unit dark matter in massive halos
of mass $M$ should be proportional to $\propto M^{-\alpha}$ with $\alpha
= 0.25$ for the PCSz, as compared to $\alpha = 0.08 $ for the LCRS. The
change of $\alpha$ from the LCRS to the PSCz is not 
unexpected, since IRAS galaxies tend to avoid high-density cluster
regions. The biased mock catalogs fit the 2PCF of the PSCz galaxies
extremely accurately. We also have selected samples from the GIF
simulation and shown that appropriate data sets can be drawn from that
numerical catalog which give similar 2PCF.

The pairwise peculiar velocity dispersion measured from the PSCz has a
much lower value, about $300 \kms$ at $r_p = 1 \mpc$, than the LCRS
result at that separation of $570 \pm 80 \kms$.  All the simulation
models which are consistent with the 2PCF of the PSCz predict
significantly larger values for PVD, although the CLW bias reduces the
PVD to within the $1 \sigma$ limit of this value, at least near $3 \mpc$
(if $\Omega_0=0.2$).  As discussed in the last section, the decrease of
$\sigma_{12}(r_p)$ for $r_p \ls 1\mpc$ for the PSCz data is significant,
and cannot be reproduced by the models.  To test the robustness of this
important result, we applied another phenomenological biasing
prescription to our simulations which mimics the observed
density-morphology relation of galaxies in clusters.  More specifically,
we have selected particles within halos with the probability equal to
the fraction of spiral galaxies in rich clusters (Dressler 1980). The
fraction is set to be independent of the halo mass for simplicity.
This biasing model does reduce both the 2PCF and PVD slightly, but the
effect is much smaller than the difference between the CLW models and
the PCSZ result. This strengthens our conclusion that the steep decline
of PVD on small scales cannot be explained with the existing CDM model.

In conclusion, we find that the infrared selected galaxies in the PSCz
have a lower and less steep two-point correlation function than the
optically selected sample of the LCRS. The PVD of the PSCz galaxies is
also lower, and has a different $r_p$ dependence than the corresponding
quantity measured from the LCRS.  Spatially-flat CDM models with
$\Omega_0 \sim 0.3$ can give a good fit to the data, except for the
small-scale behavior of the PVD, if a strong cluster antibias is
imposed. The infrared selected galaxies require a much stronger CLW bias
than the optically selected ones.

\acknowledgments 

We would like to thank G. Kauffmann for her advice on using the GIF
data, and an anonymous referee for comments.  Y.P.J. and G.B. are
grateful for the hospitality extended toward them at Department of
Physics, University of Tokyo, and for support during their stay at
RESCEU. Y.P.J. is supported in part by the One-Hundred-Talent Program,
by NKBRSF(G19990754) and by NSFC(No.10043004), and G.B. by SFB375.

%%%%%%%%%%%%%%%%%%%%%%%%%%%%%%%%%%%%%%%%%%%%%%%%%%%%%%%%%%%%%%%%%%

%%%%%%%%%%%%%%%%%%%%%%%%%%%%%%%%%%%%%%%%%%%%%%%%%%%%%%%%%%%%%%%%%%%%%%%
\begin{deluxetable}{lcccccccc}
\tablecolumns{9}
\tablewidth{0pc}
\tablecaption{Simulation model parameters.
\label{tab:simulation}}
\tablehead{
\colhead{Model} & \colhead{$\Omega_0$} &  \colhead{$\lambda_0$}
& \colhead{$\Gamma$\tablenotemark{\dagger}} & \colhead{$\sigma_8$}
& \colhead{$L_{\rm box}$[$h^{-1}$Mpc]} 
& \colhead{$\eta_{\rm grav}$[$h^{-1}$kpc]\tablenotemark{\ddagger}} 
& \colhead{$N_{\rm particle}$}
& \colhead{$m_{\rm particle}$[$M_\odot$]} 
}\startdata
LCDM & 0.3 & 0.7 & 0.2 & 1.0 & 100 &  39 & $256^3$ & $5.0\times 10^{9}$ \\
LCDM & 0.3 & 0.7 & 0.2 & 1.0 & 300 & 117 & $256^3$ & $1.3\times 10^{11}$ \\
LCDM & 0.3 & 0.7 & 0.2 & 1.0 & 800 & 400 & $300^3$ & $1.5\times 10^{12}$ \\
SCDM & 1.0 & 0.0 & 0.5 & 0.6 & 300 & 117 & $256^3$ & $4.5\times 10^{11}$ \\
GIF  & 0.3 & 0.7 & 0.21 & 0.9 & 141 & 20 & $256^3$ & $1.4 \times 10^{10}$\\
\enddata
\tablenotetext{\dagger}{Shape parameter of the power spectrum.}
\tablenotetext{\ddagger}{Gravitational softening length.}
\end{deluxetable}
%%%%%%%%%%%%%%%%%%%%%%%%%%%%%%%%%%%%%%%%%%%%%%%%%%%%%%%%%%%%%%%%%%%%%%%

\clearpage

%%%%%%%%%%%%%%%%%%%%%%%%%%%%%%%%%%%%%%%%%%%%%%%%%%%%%%%%%%%%%%%%%%%%%%%%%%%

\begin{thebibliography}{}
\bibitem[Baker, Davis, \& Lin (2000)]{bdl00} Baker, J.\ E., 
  Davis, M., \& Lin, H.\ 2000, \apj, 536, 112 
\bibitem[Bardeen et al.(1986)] {bbks86} Bardeen, J., Bond, J.R., Kaiser, N.,
  \& Szalay, A.S., 1986, {ApJ}, {304}, 15
\bibitem[Beichman et al.(1988)]{es88} Beichman, C.A. et al. 1988, 
  IRAS Catalogs and Atlases, Vol.1: Explanatory Supplement (JPL)
\bibitem[Carlberg et al.(1996)]{carl96} Carlberg, R.,G., Yee, H.,K.,C.,
   Ellingson, E., Abraham, R., Gravel, P., Morris, S., \& Pritchet, C.J., 1996,
   ApJ, 462, 32
\bibitem[Davis \& Peebles(1983)] {dp83} 
   Davis, M., \& Peebles, P.J.E., 1983, ApJ, 267, 465
\bibitem[Diaferio \& Geller(1996)] {diaferio96} 
   Diaferio, A., \& Geller, M.J. 1996, ApJ, 467, 19
\bibitem[Dressler (1980)] {dressler} 
  Dressler, A. 1980, ApJ, 236, 351
\bibitem[Efstathiou et al.(1988)] {efstathiou}
   Efstathiou, G., Frenk, C. S., White, S. D. M., \& Davis, M. 1988, 
   MNRAS, 235, 715
\bibitem[Magira, Jing, \& Suto (1999)] {magira}
    Magira, H., Jing, Y.P., \& Suto, Y. 2000, 528,30
\bibitem[Fisher et al.(1994a)] {fisher94a} Fisher, K.B., Davis, M., 
   Strauss, M.A., Yahil, A., \& Huchra, J.P. 1994a, MNRAS, 267, 927
\bibitem[Fisher et al.(1994b)]{fisher94b} Fisher, K.\ B., Davis, M., 
   Strauss, M.\ A., Yahil, A., \& Huchra, J.\ 1994b, \mnras, 266, 50 
\bibitem[Hamilton \& Tegmark(2001)] {hamilton01} Hamilton, A.J.S. \&
    Tegmark, M.. 2001, MNRAS, in press (astro-ph/0008392)
\bibitem[Jing(1998)] {jing98} Jing, Y.P. 1998, ApJ, 503, L9
\bibitem[Jing, Mo, \& B\"orner(1998)]{jmb98} Jing, Y.P., Mo, H.J., \&
   B\"orner, G. 1998, ApJ, 494, 1 (JMB98)
\bibitem[Jing \& B\"orner(1998)]{jb98a} Jing, Y.P., \& B\"orner, G.
  1998, ApJ, 503, 37
\bibitem[Jing \& B\"orner(2001)] {jb01} Jing, Y.P., \& B\"orner, G.
   2001, MNRAS (in press); astro-ph/0101211
\bibitem[Jing \& Suto(1998)] {jingsuto98} Jing, Y.P., \& Suto, Y. 1998,
   ApJ, 494, L5
\bibitem[Juszkiewicz et al.(2000)]{roman00}Juszkiewicz, R., 
Ferreira, P. G., Feldman, H. A., Jaffe, A. H., \& Davis, M. 2000,
Science, 287, 109
\bibitem[Kauffmann et al.(1999a)] {gif99a} Kauffmann, G., Colberg, J.M.,
   Diaferio, A., \& White, S.D.M. 1999, MNRAS, 303, 188
\bibitem[Kauffmann et al.(1999b)] {gif99b} Kauffmann, G., Colberg, J.M.,
   Diaferio, A., \& White, S.D.M. 1999, MNRAS, 307, 529
\bibitem[Mo, Jing, \& B\"orner(1992)]{mo92} Mo, H.J., Jing, Y.P.,
   \& B\"orner, G., 1992, ApJ, 392, 452
\bibitem[Mo, Jing, \& B\"orner(1993)]{mo93} Mo, H.J., Jing, Y.P.,
   \&  B\"orner, G., 1993, MNRAS, 264, 825
\bibitem[Mo, Jing, \& B\"orner(1997)]{mo97} 
   Mo, H.J., Jing, Y.P., \& B\"orner, G., 1997, MNRAS, 286, 979
\bibitem[Nolthenius, Klypin, \& Primack(1997)]{nkp97} 
 Nolthenius, R., Klypin, A.\ A., \& Primack, J.\ R.\ 1997, \apj, 480, 43 
\bibitem[Ostriker \& Suto(1990)]{os90} 
 Ostriker, J.P. \& Suto, Y.\ 1990, \apj, 348, 378
\bibitem[Peacock \& Smith(2000)] {pea2000} 
  Peacock, J.A., \& Smith, R.E. 2000, MNRAS, 318, 1144
\bibitem[Peebles(1995)]{p95} Peebles, P.\ J.\ E.\ 1995, \apj, 449, 52 
\bibitem[Saunders et al.(2000)] {saunders00} Saunders, W. et al. 2000,
  MNRAS, 317, 55
\bibitem[Schlegel et al.(1994)]{localgroup} 
   Schlegel, D., Davis, M., Summers, F., \& Holtzman, J.\ A.\ 1994, 
   \apj, 427, 527 
\bibitem[Seaborne et al.(1999)] {seaborne99} Seaborne, M.D. et al. 1999,
  MNRAS, 309, 89
\bibitem[Seljak(2000)] {seljak00} Seljak, U. 2000, MNRAS, 318, 203
\bibitem[Seto \& Yokoyama(1998)] {seto98} Seto, N., \& Yokoyama, J.I.
   1998, ApJ, 492, 421
\bibitem[Sheth(1996)] {sheth96} Sheth, R.K. 1996, MNRAS, 279, 1310
\bibitem[Sheth et al.(2000)] {sheth00} Sheth, R.K., Lam, H., Diaferio, A.,
  \& Scoccimarro, R. 2000, MNRAS, in press (astro-ph/0009167)
\bibitem[Spergel \& Steinhardt(2000)] {ss00}
   Spergel, D.N., \& Steinhardt, P.J. 2000, Phys.Rev.Lett., 84, 3760 
\bibitem[Strauss \& Willick(1995)]{sw95} 
   Strauss, M.A., \& Willick, J. A. 1995, \physrep , 261, 271
\bibitem[Strauss, Ostriker, \& Cen(1998)]{straussetal1998} 
   Strauss, M.\ A., Ostriker, J.\ P., \& Cen, R.\ 1998, \apj, 494, 20 
\bibitem[Suto \& Jing(1997)]{sutojing97} 
   Suto Y., \& Jing Y.P. 1997, ApJS, 110, 167
\bibitem[Szapudi et al.(2000)]{szapudi} Szapudi, I., Branchini, E., 
   Frenk, C.\ S., Maddox, S., \& Saunders, W.\ 2000, \mnras, 318, L45 
\bibitem[Yoshida et al.(2000)]{yoshida} Yoshida, N., Springel, V., 
   White, S.D.M., \& Tormen, G. 2000, ApJ, 544, 87
\end{thebibliography}
\end{document}